\documentclass[preprint,preprintnumbers,nofootinbib]{revtex4}
\usepackage{graphicx}

\newcommand{\ev}{{\rm eV}}
\newcommand{\mev}{{\rm MeV}}
\newcommand{\gev}{{\rm GeV}}

\begin{document}

\title{Constraints on the Timeon Model}
\author{Takeshi Araki and C.~Q.~Geng}
\affiliation{Department of Physics, National Tsing Hua University,
Hsinchu, Taiwan 300.}

\begin{abstract}
The timeon model recently proposed by Friedberg and Lee has a potential problem
 of flavor changing neutral currents (FCNCs) if the mass of the timeon is small.
In order to avoid,
% the problem without fine-tuning, 
we introduce a small
 dimensionless parameter to suppress FCNCs.
Even in this case, we find that the timeon mass must be larger than $151\ \gev$ to satisfy
all the constraints from processes involving FCNCs in the 
 quark  sectors. 
We also extend the timeon model to the lepton sector and examine the leptonic processes.
%.
\end{abstract}

\maketitle

\section{Introduction}
$CP$ violation (CPV) is one of the most important features in particle physics.
In the standard model (SM),
it arises from a unique physical phase of the Cabibbo-Kobayashi-Maskawa (CKM)
quark mixing matrix \cite{ckm}.
However, the origin of CPV remains unclear. 
On the other hand, it is unknown whether CPV also occurs in the lepton sector.

In general, there are two possible sources of CPV: complex  couplings of the Yukawa interactions and 
complex vacuum expectation values (VEVs) of the scalar fields,
which explicitly and spontaneously break $CP$ symmetry, such as the cases of the SM and 
multi-Higgs doublet models~ \cite{CPV},
respectively. 

Recently, a new spontaneous $CP$ violation mechanism has been
proposed by Friedberg and Lee \cite{timeon}.
They introduced a  gauge singlet pseudoscalar which couples to quarks 
to yield a new type of Yukawa interactions.
As the Yukawa couplings
 are assumed to be real, the system is $CP$ and $T$ conserving.
Once the pseudoscalar acquires a VEV,
both $CP$ and $T$ are spontaneously broken.
The excitation of the pseudoscalar from its vacuum predicts a new neutral
 scalar field named as ``timeon''.
One of the most interesting features of the proposal is that the pseudoscalar is also
 responsible for the masses of up and down quarks, 
 but not for the heavy quarks.
As a result, the timeon mass and the VEV can be much lower than electroweak scale.

The timeon model, however, has a potential problem of flavor changing neutral
 currents (FCNCs) mediated by the timeon.
As we shall explain later, the timeon mass is strongly constrained by the mass
 mixings of the neutral mesons and must be much larger than its VEV.
This fact implies the strongly interacting quartic timeon coupling and
the broken down of the perturbation.
In this paper, in order to save the model,
% without fine-tuning, 
we introduce a
 small parameter $\epsilon$ and
 re-define the VEV as $\tau_0 \rightarrow \epsilon \tau_0$. 
We will demonstrate that the small $\epsilon$ can relax the
phenomenological constraints, which permit
a much smaller mass of  the timeon.
We would also extend the timeon model to the lepton sector and
explore leptonic FCNC processes, such as the lepton number violating decays and 
the muon anomalous magnetic moment.

The paper is organized as follows.
In Sec. 2, we give a brief review of the timeon model and calculate the 
 mixing parameters in the neutral meson systems.
In Sec. 3, we extend the timeon model to the lepton sector.
We present the summary in Sec. 4.

\section{Quark sector}
\subsection{Model}
In the theory of timeon, proposed by
Friedberg and Lee~\cite{timeon},
the quark mass Hamiltonian is written as
\begin{eqnarray}
{\cal H}_q 
 &=&\bar{u}_i[G_u + i\gamma_5  \tau_0 F]_{ij}u_j
        +\bar{d}_i[G_d + i\gamma_5  \tau_0 F]_{ij}d_j \nonumber\\
 &=&\bar{u}_{Li}[G_u + i \tau_0 F]_{ij}u_{Rj}
        +\bar{d}_{Li}[G_d + i \tau_0 F]_{ij}d_{Rj}
        +h.c.\ ,\label{eq:quark}
\end{eqnarray}
where the constant $\tau_0$ is the VEV of
the new $CP$ odd and $T$ odd,
gauge singlet pseudoscalar field $\tau(x)$, $i,j=1,\cdots, 3$ stand for family indices and
$G_{u,d}$ and $F$ are $3 \times 3$ matrices, given by
\begin{eqnarray}
G_{u,d}=
 \left(\begin{array}{ccc}
 b_{u,d} \eta_{u,d}^2 (1+\xi_{u,d}^2) & -b_{u,d} \eta_{u,d} &
 -b_{u,d} \xi_{u,d} \eta_ {u,d}\\
 -b_{u,d} \eta_{u,d} & b_ {u,d}+ a_{u,d} \xi_{u,d}^2 & -a_{u,d} \xi_{u,d} \\
 -b_{u,d} \xi_{u,d} \eta_{u,d} & -a_{u,d} \xi_{u,d} & a_{u,d} + b_{u,d}
 \end{array}\right)\label{eq:Gq}\ 
\end{eqnarray}
and
\begin{eqnarray}
F=
 \left(\begin{array}{ccc}
 \cos^2\alpha_q & \sin\alpha_q \cos\alpha_q \cos\beta_q &
  \sin\alpha_q \cos\alpha_q \sin\beta_q \\
 \sin\alpha_q \cos\alpha_q \cos\beta_q & \sin^2\alpha_q \cos^2\beta_q &
  \sin^2\alpha_q \sin\beta_q \cos\beta_q \\
 \sin\alpha_q \cos\alpha_q \sin\beta_q & \sin^2\alpha_q \sin\beta_q \cos\beta_q &
  \sin^2\alpha_q \sin^2\beta_q 
 \end{array}\right)\ \label{eq:Fmatrix}
\end{eqnarray}
with the real parameters of $a_{u,d} , b_{u,d} , \eta_{u,d}$ and $\xi_{u,d}$ and two angles of
$\alpha_q$ and $\beta_q$, respectively.
Furthermore,  $\eta_{u,d}$ and $\xi_{u,d}$ in Eq. (\ref{eq:Gq}) can be also defined by four angles as
\begin{eqnarray}
&&\xi_u = \tan\phi_u,\ \eta_u = \tan\varphi_u \cos\phi_u\ , \\
&&\xi_d = \tan\phi_d,\ \eta_d = -\tan\varphi_d \cos\phi_d\ .
\end{eqnarray}
The potential of the pseudoscalar field is given by~\cite{timeon}
\begin{eqnarray}
V(\tau)&=& -{1\over 2}\lambda\tau^2\left(\tau_0^2-{1\over 2}\tau^2\right)\ ,
%<\tau (x)>=\tau_0 + t(x)\ ,\label{eq:vev}
\end{eqnarray}
which leads to the mass of the new quantum, $i.e.$, timeon, to be
\begin{eqnarray}
M_T&=&(2\lambda)^{1\over 2}\tau_0\,.
%\tau_0 + t(x)\ ,
\label{eq:MT}
\end{eqnarray}

As  seen from Eq. (\ref{eq:Gq}), $G_{u,d}^* =G_{u,d}$ and $\det G_{u,d}=0$, and thus
 up and down quarks of the first generation  are massless before $\tau(x)$ receives the VEV.
After $\tau(x)$ gets the VEV, $CP$ and $T$ are spontaneously broken
and the first generation quarks acquire small masses.
Note that since the timeon is not responsible for the masses of heavy quarks,
its VEV can be much lower than the electroweak scale. 

\subsection{Flavor Changing Neutral Currents}
In Ref. \cite{timeon}, it was estimated that
 $\tau_0 \simeq 33\ \mev$, which implies that 
 \begin{eqnarray}
 M_T&\leq& 47\ \mev
 \label{Eq:MT0}
 \end{eqnarray}
 from Eq. (\ref{eq:MT}) if
 $\lambda\leq 1$.
 Therefore, the mass of the timeon is much lower than
the electroweak scale.
However, a light timeon could give rise to dangerous
 FCNC processes because in general $G_{u(d)}$ and $F$
 cannot be diagonalized simultaneously.
In fact, in the mass eigenstate basis of quarks, flavor changing timeon
 couplings are given by
\begin{eqnarray}
 &&F^u \equiv U_{u}^{T}FU_{u}=
  \left(\begin{array}{ccc}
   0.08 & -0.14 & 0.23 \\
   -0.14 & 0.25 & -0.41 \\
   0.23 & -0.41 & 0.67
  \end{array}\right)\ , \nonumber\\
 &&F^d \equiv U_{d}^{T}FU_{d}=
    \left(\begin{array}{ccc}
   0.15 & -0.18 & 0.31 \\
   -0.18 & 0.21 & -0.36 \\
   0.31 & -0.36 & 0.64
  \end{array}\right)\ ,
  \label{Eq:FU}
\end{eqnarray}
where $U_u$ and $U_d$ are the unitary matrices to diagonalize  up and down-type
 quarks, respectively.
\begin{table}[t]
\begin{center}
\begin{tabular}{|c|c||c|c|}\hline
 parameter & input & parameter &input \\ \hline
 $m_u(2\ \gev)$ & $2.5\times 10^{-3}\ \gev$ & $f_K$ & $0.16\ \gev$ \\ \hline
 $m_d(2\ \gev)$ & $5\times 10^{-3}\ \gev$ & $f_{B_s}$ & $0.24\ \gev$ \\ \hline
 $m_s(2\ \gev)$ & $0.095\ \gev$ & $f_{B_d}$ & $0.198\ \gev$ \\ \hline
 $m_c(m_c)$ & $1.25\ \gev$ & $f_D$ & $0.223\ \gev$ \\ \hline
 $m_b(m_b)$ & $4.2\ \gev$ & $M_K$ & $0.497\ \gev$ \\ \hline
 $M_{B_s}$ & $5.366\ \gev$ & $M_{B_d}$ & $5.280\ \gev$ \\ \hline
 $M_D$ & $1.865\ \gev$ &  &  \\ \hline
\end{tabular}
\end{center}
\caption{Values of parameters used in the calculation.
Here, we have used the same quark masses as those in Ref. \cite{timeon},
while the decay constants and the meson masses  are the center values  in
 Ref. \cite{kifune}.}
\label{tab:para}
\end{table}
From the values in Eq. (\ref{Eq:FU})  and inputs in Table \ref{tab:para}, 
the mass mixing parameter  in the neutral $K$ meson system due to the timeon contribution at tree level is
estimated to be
\begin{eqnarray}
 \Delta M_K^{timeon} 
 &=&2\left| \frac{2F_{ds}^d (F_{sd}^d)^{*} }{M_T^2}
 <K^0|\bar{d}_L^\alpha s_R^\alpha \bar{d}_R^\beta s_L^\beta |\bar{K^0}>
 \right.\nonumber\\
 &&\hspace{1cm}\left. +\frac{(F_{ds}^d)^2 + (F_{sd}^d)^{*2} }{M_T^2}
 <K^0|\bar{d}_L^\alpha s_R^\alpha \bar{d}_L^\beta s_R^\beta |\bar{K^0}>
 \right| \nonumber\\
 &\sim & \frac{1.76\times 10^{-3}\ \gev^3}{M_T^2}\,.
% \nonumber\ ,\\
 %&<& \Delta M_K^{exp}\simeq 3.5\times 10^{-15}\ \gev\ ,
 \label{eq:mk}
\end{eqnarray}
%where $M_T$ is the mass of the timeon.
% and matrix elements are given by \cite{masiero}
%\begin{eqnarray}
%<K^0|\bar{d}_L^\alpha s_R^\alpha \bar{d}_R^\beta s_L^\beta |\bar{K^0}>
%&=&\left[ \frac{1}{24}+\frac{1}{4}\left(\frac{M_K}{m_s + m_d}\right)^2 \right]
%        M_K f_K^2\nonumber\ ,\\
%&=& 0.0791\ \gev^3\ ,
%\end{eqnarray}
%\begin{eqnarray}
%<K^0|\bar{d}_L^\alpha s_R^\alpha \bar{d}_L^\beta s_R^\beta |\bar{K^0}>
%&=&-\frac{5}{24}\left(\frac{M_K}{m_s + m_d}\right)^2
%        M_K f_K^2\nonumber\ ,\\
%&=& -0.0655\ \gev^3\ .
%\end{eqnarray}
From the experimental data of 
$\Delta M_K^{exp}\simeq 3.5\times 10^{-15}\ \gev$, 
without loss of generality we find the lower limit of the timeon mass as 
%$M_T \ge 0.7\times 10^{6}\ \gev$, 
\begin{eqnarray}
M_T \ge 7\times 10^{5}\ \gev\,,
 \label{Eq:MT1}
 \end{eqnarray}
which is much higher than that in Eq. (\ref{Eq:MT0}).
Here, we have ignored the renormalization group effects as well as some hadronic uncertainties.
We note that the mixing parameters  in other neutral meson systems also lead to similar constraints on $M_T$.
Therefore,
in the timeon model of Ref. \cite{timeon}, $M_T\gg \tau_0$,
which requires the strongly interacting quartic timeon coupling, $i.e.$, $\lambda\gg 1$.
%, and the  broken down of the perturbation.
%If we fine-tune $\alpha_q$ in Eq. (\ref{eq:Fmatrix}) to $\pi/2$, we may be able
 %to explain more large V.E.V.
%But we do not consider such a situation in this paper.

In order to solve the problem, in this paper, we introduce a small dimensionless
 parameter $\epsilon$ for the terms related to $F$ in Eq. (\ref{eq:quark}), which is equivalent
to a  re-defined  VEV  of
\begin{eqnarray}
 \tau_0 \rightarrow \epsilon\tau_0 \simeq 33\ \mev \,.\label{eq:vev2}
\end{eqnarray}
If $\epsilon$ is very small, 
the constraints from the FCNC processes can be relaxed
becouse the left hand side of Eq. (\ref{Eq:MT1}) is replaced by 
$M_T/\epsilon$.
%, while $\tau_0$ can be large.
In this case,
the constraint from $\Delta M_K$ becomes
\begin{eqnarray}
 M_T \ge 151\ \gev\ .
\end{eqnarray}
%In this paper, b
By assuming that $\tau_0\simeq M_T\sim 151\ \gev$, $i.e.$, $\lambda\simeq 1/2$,
%then
we  obtain that $\epsilon \sim 0.22\times 10^{-3}$ from
 Eq. (\ref{eq:vev2}).
 %as well, and this indicates 
%$\epsilon \simeq 0.22\times 10^{-3}$.
However, If we use a  smaller $\tau_0$ or a larger $\epsilon$, 
$M_T$ has to be larger
 than $151\ \gev$.
In this paper, 
%we take $M_T\sim \tau_0$ and take $151\ \gev$
% as the lowest value of the timeon mass.
%Hence 
we take $151\ \gev$ as the lowest value of the timeon mass.
%in this paper.
Consequently, we can also estimate the mixings 
in $B_s$, $B_d$ and $D$ meson systems from the timeon contributions.
By using  $M_T \sim 151\ \gev$, we obtain
\begin{eqnarray}
 &&\Delta M_{B_s}^{timeon} \sim 0.365\times 10^{-13}\ \gev 
   <\Delta M_{B_s}^{exp}\simeq 1.17\times 10^{-11}\ \gev\ ,\\
 &&\Delta M_{B_d}^{timeon} \sim 0.178\times 10^{-13}\ \gev 
   <\Delta M_{B_d}^{exp}\simeq 3.34\times 10^{-13}\ \gev\ ,\\
 &&\Delta M_{D}^{timeon} \sim 0.205\times 10^{-14}\ \gev
   <\Delta M_{D}^{exp}\simeq 1.4\times 10^{-14}\ \gev\ .
\end{eqnarray}
It is clear that all the FCNC processes are suppressed if $\epsilon\le 0.22\times 10^{-3}$
and $M_T \ge 151\ \gev$.

Now we would like to briefly comment on the possible origin of the small parameter
 $\epsilon$.
Since $\tau(x)$ is a gauge singlet scalar field, 
it cannot be
 incorporated into the renormalizable $SU(2)_L \times U(1)_Y$ model.
One way out is to consider
a non-renormalizable interacting term, such as,
\begin{eqnarray}
 \frac{F_{ij}}{\Lambda}\bar{\psi}_{Li} \Phi \psi_{Rj} \tau\ ,
\end{eqnarray}
where $\Phi$ is a $SU(2)_L$ doublet scalar and
 $\Lambda$ is a typical energy scale of the model.
In this case, $<\Phi>/\Lambda$ might be able to explain the existence of 
the small parameter $\epsilon$.
As the construction of realistic models goes beyond our purpose of 
this paper, we do not discuss this point further.

\section{Lepton sector}
\subsection{Model}
In this section, we extend the timeon model to the lepton sector.
The lepton mass Hamiltonian is similar to Eq. (\ref{eq:quark}),
 given by
\begin{eqnarray}
{\cal H}_l
 &=&\bar{\ell}_{i}[G_\ell + i\gamma_5 \epsilon \tau_0 F]_{ij}\ell_{j}
        +\bar{\nu}_i^{(c)}[G_\nu + i\gamma_5 \epsilon_\nu \tau_0 F]_{ij}\nu_j\ ,
         \label{eq:lepton}
\end{eqnarray}
where $\ell_i$ and $\nu_i$ represent charged leptons and neutrinos, respectively.
Note that in the charged lepton sector, we use the same parameter $\epsilon$ as that of
the quark sector, whereas in the
neutrino sector we assume a different small parameter $\epsilon_\nu$ as
 neutrino masses would be due to some new physics such as the
See-Saw mechanism, which may have a different origin from that of the charged fermions
\footnote{We do not restrict ourselves to the case of Majorana neutrinos.}.
The matrix $F$ takes the same form as Eq. (\ref{eq:Fmatrix}) but described
by different angles $\alpha_l$ and $\beta_l$.
For  $G_{\ell,\nu}$, we assume the following simple textures of
\begin{eqnarray}
G_\ell=
 \left(\begin{array}{ccc}
 0 & 0 & 0 \\
 0 & b_\ell + a_\ell & a_\ell \\
 0 & a_\ell & a_\ell + b_\ell
 \end{array}\right)\label{eq:Gell}\ ,
\end{eqnarray}
\begin{eqnarray}
G_\nu=
 \left(\begin{array}{ccc}
 \frac{1}{2}b_\nu & \sqrt{\frac{1}{2}}b_\nu & 0 \\
 \sqrt{\frac{1}{2}}b_\nu &  b_\nu & 0 \\
 0 & 0 & a_\nu + b_\nu
 \end{array}\right)\label{eq:Gnu}\,,
\end{eqnarray}
respectively.
Note that this corresponds to
$\eta_\ell = \xi_\nu = 0,\xi_\ell=-1$ and $\eta_\nu = -\sqrt{\frac{1}{2}}$
\cite{FL}.

Here, we would like to comment on several features of the model.
As shown in Ref. \cite{FL}, Eqs. (\ref{eq:Gell}) and (\ref{eq:Gnu})
imply the exact tri-bimaximal lepton mixing matrix \cite{TB}.
That is,
\begin{eqnarray}
 \sin^2 \theta_{12}= \frac{1}{3}\ ,\ \sin^2 \theta_{23} = \frac{1}{2}\ ,\ 
 \sin^2 \theta_{13} =0\ .
\end{eqnarray}
After $\tau(x)$ receives the VEV,  the mixing matrix deviates from exact tri-bimaximal
mixing.
However, if $\alpha_l =0$, there is no flavor changing timeon coupling in the charged
 lepton sector, while $\sin^2 \theta_{13}=0$ and
$\sin^2 \theta_{23}=1/2$ remain.
On the other hand, 
if $\alpha_l\neq 0$, $\sin^2 \theta_{13}$ and $\sin^2 \theta_{23}$ will be different from
$0$ and $1/2$, respectively, and flavor changing timeon couplings in the charged
 lepton sector are induced.
As in the case of the quark sector, $CP$ is spontaneously broken.
We note that if $\alpha_l =0$, the $CP$ violating Dirac phase in the neutrino sector will not show up 
due to $\sin^2 \theta_{13}=0$.
%.

\begin{figure}[t]
\begin{center}
\includegraphics*[width=0.8\textwidth]{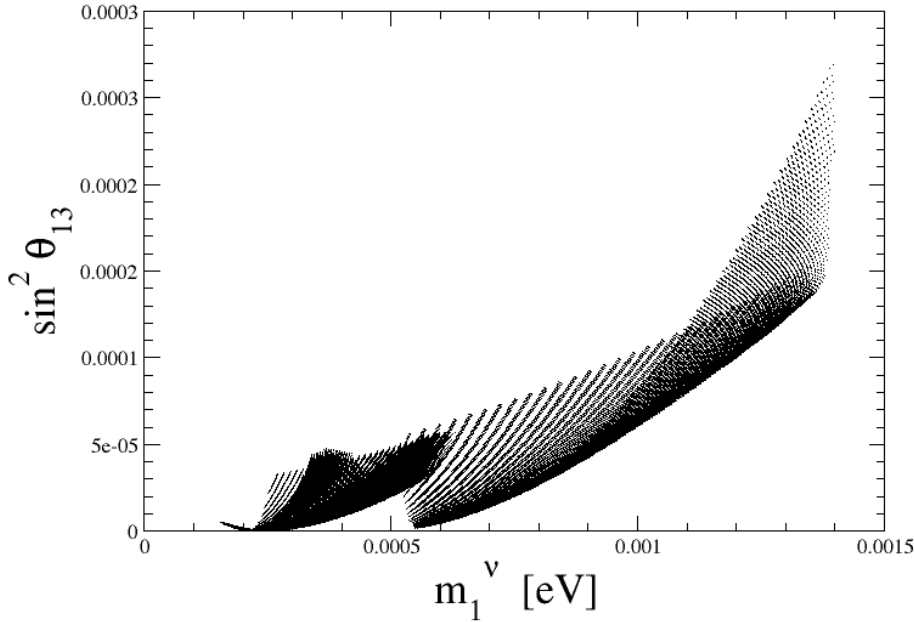}
\caption{\footnotesize
The predictions of the model in $m_1^\nu - \sin^2 \theta_{13}$ plane.
}\label{fig:m1-rct}
\end{center}
\end{figure}

Since our model is a special case of Ref. \cite{timeon},
the analytic study 
is the same as that in Ref. \cite{timeon}.
Because the values of $\epsilon$ and $\tau_0$ are already determined
as $\epsilon=0.22\times 10^{-3}$ and $\tau_0 = 151\ \gev$ in the quark sector,
 the model is described by seven 
 parameters: $a_\ell , b_\ell , a_\nu , b_\nu , \alpha_l , \beta_l$ and $\epsilon_\nu$,
which can be  fixed by  seven physical quantities.
In our numerical calculation, we take~\cite{PDG}
\begin{eqnarray}
 m_e = 0.511\ \mev ,\ m_\mu = 105.658\ \mev ,\ 
 m_\tau = 1776.84 \pm 0.17\ \mev ,
 \label{Eq:D1}
\end{eqnarray}
%from Particle Data Group (PDG) \cite{PDG} 
and~\cite{osi}
\begin{eqnarray}
 \Delta m_{21}&=&(7.45- 7.88)\times 10^{-5}\ \ev^2 \,,\
 \Delta m_{31}=(2.29-2.52)\times 10^{-3}\ \ev^2 , 
 \nonumber\\
 \sin^2 \theta_{12}&=& 0.288- 0.326 \,,\
 \sin^2 \theta_{23}= 0.44- 0.57 \,.
% \sin^2 \theta_{13}> 0.026 ,
 \label{Eq:D2}
\end{eqnarray}
%best fit values in Ref. \cite{osi}.
One may think that if we also use $\sin^2 \theta_{13}$, we can determine the
 value of $\epsilon\tau_0$ without the result in
 the quark sector.
However, the experimental upper limit of $\sin^2 \theta_{13}$ is too mild to constrain
 the model.
 % .
Nevertheless, the value of $\sin^2 \theta_{13}$ is obtained as a prediction
 of the model.
 From the  quantities in Eqs. (\ref{Eq:D1}) and  (\ref{Eq:D2}), we get
$\alpha_l \simeq 82.89^\circ$.
Hence, our model gives a small  $\sin^2 \theta_{13}$ as well as FCNCs
 mediated by the timeon.
In Figure \ref{fig:m1-rct}, we  show the allowed values of
$m_1^\nu$ and $  \sin^2 \theta_{13}$.

\subsection{Leptonic Processes}
As  mentioned in the previous subsection, $\alpha_l \neq 0$ results in the
existence of flavor changing timeon couplings.
The branching ratios of the lepton flavor violating (LFV)
 decays due to these coupling are given by~\cite{lfv}
\begin{eqnarray}
&&Br(\mu\rightarrow e\gamma)
=\frac{\alpha_{em}\tau_\mu}{2^{10}\pi^4}
  \frac{m_\mu^3 m_\tau^2}{M_T^4}
  \left(\epsilon F_{\mu\tau}\ \epsilon F_{e\tau}\right)^2
  \left| \ln\frac{m_\tau^2}{M_T^2} + \frac{3}{2} \right|^2\ ,
  \label{eq:meg}\\
  &&Br(\ell^- \rightarrow \ell_3^- \ell_2^+ \ell_1^-)
=\frac{5}{3}\frac{\tau_\ell}{2^{11}\pi^3}\frac{m_\ell^5}{M_T^4}
  \left(\epsilon F_{\ell_3\ell}\ \epsilon F_{\ell_2\ell_1}\right)^2\,,
%&&Br(\mu^- \rightarrow e^- e^+ e^-)
%=\frac{5}{3}\frac{\tau_\mu}{2^{11}\pi^3}\frac{m_\mu^5}{M_T^4}
%  \left(\epsilon F_{e\mu}\ \epsilon F_{ee}\right)^2\ ,\\
%&&Br(\tau^- \rightarrow e^- e^+ e^-)
%=\frac{5}{3}\frac{\tau_\tau}{2^{11}\pi^3}\frac{m_\tau^5}{M_T^4}
%  \left(\epsilon F_{e\tau}\ \epsilon F_{ee}\right)^2\ ,\\
%&&Br(\tau^- \rightarrow \mu^- \mu^+ \mu^-)
%=\frac{5}{3}\frac{\tau_\tau}{2^{11}\pi^3}\frac{m_\tau^5}{M_T^4}
%  \left(\epsilon F_{\mu\tau}\ \epsilon F_{\mu\mu}\right)^2\ ,
\label{eq:t3m}
\end{eqnarray}
where $\alpha_{em}$ is the fine structure constant, $\ell=\mu$ or $\tau$, $\ell_i=e$ or $\mu$, and $\tau_{\ell}$
is the life time of the $\ell$ lepton.
From Eqs. (\ref{eq:meg}) and (\ref{eq:t3m}), we obtain
\begin{eqnarray}
&&Br(\mu\rightarrow e\gamma)
\simeq 10^{-16} - 10^{-21}\,,
\nonumber\\
&&Br(\mu^- \rightarrow e^- e^+ e^-)
\simeq 10^{-21} - 10^{-26}\,,
\nonumber\\
&&Br(\tau^- \rightarrow e^- e^+ e^-)
\simeq 3\times 10^{-21}\,,
\nonumber\\
&&Br(\tau^- \rightarrow \mu^- \mu^+ \mu^-)
\simeq 10^{-19} - 10^{-35}\ .
\end{eqnarray}
Similarly, we find that the timeon contribution to 
the muon anomalous magnetic moment is
\begin{eqnarray}
 \Delta a_\mu &=&\frac{1}{8\pi^2}
 \frac{m_\mu^2 m_\tau^2}{M_T^2}
 \left(\epsilon F_{\mu\tau}\right)^2
  \left( \ln\frac{m_\tau^2}{M_T^2} + \frac{3}{2} \right)\nonumber\\
 &\simeq& 10^{-16} - 10^{-21}\ .
\end{eqnarray}
It is clear that all contributions to the leptonic processes from the timeon are too small to be measured in the
 near future.

\section{Summary}
We have investigated the timeon model, which is a new kind of
 spontaneous $CP$ violation mechanism, recently proposed by Friedberg and Lee~\cite{timeon}.
However, the original model has a potential problem of FCNCs mediated by the timeon
 if the timeon mass $M_T$ is smaller than $7\times 10^5\ \gev$.
In order to avoid  the problem, we have introduced a small dimensionless parameter 
to suppress FCNCs and thus allow a smaller $M_T$.
 Due to the constraints from the mass mixing parameters in the $K^0-\bar{K}^0$ and other neutral meson systems, 
 we have shown that  $M_T \ge 151\ \gev$.
We have also extended the timeon model to the lepton sector.
We have found  that our simple model predicts a small but non-zero $\sin^2 \theta_{13}$ as well as
 non-zero lepton flavor violating timeon couplings.
However, all contributions to the LFV decays and muon anomalous magnetic moment due to the timeon
  are not measurable. \\
 
 \noindent {\bf Acknowledgments}

This work is supported in part by
the National Science Council of ROC under Grant No:
 NSC-95-2112-M-007-059-MY3.

\end{document}